# From Discrete Trailing Returns to a Continuous Graphical Profile: Return-to-Present Curves

Lei Liu
Washington University in St. Louis

**Abstract.** Investment performance is commonly presented either as a conventional cumulative-return chart, which fixes a historical starting date and traces performance forward, or as a trailing-return table, which fixes the current endpoint but reports only a small set of prespecified horizons. These two displays have complementary limitations: fixed-start comparisons are conditional on the selected origin, whereas trailing returns provide only discrete snapshots of the underlying fixed-endpoint return function. We present the Return-to-Present (RTP) curve as a continuous fixed-endpoint representation that brings these perspectives together by holding the evaluation date fixed while allowing the hypothetical historical purchase date to vary over the available history. Familiar 1-month, 3-month, 6-month, 1-year, and longer trailing returns therefore become selected points on a continuous curve. When multiple investments are overlaid, RTP directly displays entry-date sensitivity, persistent relative advantage, crossings, and the timing and magnitude of separation without requiring selection of a single historical origin. The same endpoint-based construction naturally accommodates investments with unequal inception dates, recurring purchases, and retrospective portfolio rotation decisions in which sale and replacement-purchase dates may differ. We illustrate these uses with real investment data and discuss its relationship to momentum. RTP does not define a new return measure; its contribution is a simple graphical organization of familiar realized returns for historical comparison and decision support rather than prediction or statistical inference.

**Keywords:** investment performance visualization; fixed-endpoint returns; trailing returns; continuous return curve; momentum; historical performance profile; entry-date sensitivity; endpoint sensitivity; dollar-cost averaging; decision support

## 1. Introduction

Investment performance is commonly presented in two complementary ways. One is a conventional cumulative-return chart, in which multiple investments are normalized at a common historical starting date and their subsequent performance is plotted forward through time. This provides a continuous graphical view of performance, but the comparison is conditional on the selected starting date. A chart beginning six months ago, one year ago, or three years ago may produce different relative performance patterns, and there is often no natural reason to regard one historical date as the uniquely appropriate origin for comparison.

A second common approach is the trailing-return table, which reports returns to the present over several prespecified horizons, such as 1 month, 3 months, 6 months, 1 year, 3 years, and 5 years. Unlike the conventional cumulative-return chart, these measures share a common endpoint—the present—but they provide only a discrete set of observations from the underlying fixed-endpoint return function. Consequently, changes occurring between the reported horizons, including crossings, shifts in relative ranking, and the timing and magnitude of emerging performance differences, may not be apparent.

We propose the Return-to-Present (RTP) curve as a continuous fixed-endpoint representation that holds the evaluation date fixed while allowing the hypothetical historical purchase date to vary over the available history. RTP brings together the complementary strengths of these two familiar displays. It retains the common-endpoint perspective of trailing returns while providing the continuity of a graphical performance curve. Rather than fixing one historical starting date and tracing performance forward, RTP fixes the evaluation date—typically the present—and looks backward across all available historical purchase dates. It can therefore be viewed as a graphical generalization of the familiar trailing-return table, replacing a small set of prespecified horizons with a continuous fixed-endpoint performance profile. When multiple investments are overlaid, their relative performance, crossings, persistence, and magnitude of separation become directly visible without requiring the analyst to choose a single historical origin or reconstruct the pattern from isolated return values.

We emphasize that RTP does not define a new return quantity. Its contribution is representational rather than algebraic: RTP organizes familiar fixed-endpoint returns as a continuous graphical profile over historical purchase dates for direct multi-asset comparison. It can therefore be viewed as a graphical generalization of familiar trailing-return summaries rather than as a new return formula. The paper first defines the RTP curve and explains how to read it, then illustrates its use with real investment examples and discusses its interpretation, scope, and relationship to momentum.

## 2. Return-to-Present Curves

Let $P(t)$ denote the adjusted price of an investment at the evaluation date $t$. For a look-back horizon $h \geq 0$, define the Return-to-Present curve as

$$R(h;\, t) = \frac{P(t)}{P(t-h)} - 1$$

At each value of $h$, the curve answers a direct question: if the investment had been purchased $h$ days before the evaluation date and held through $t$, what percentage return would have been realized? At $h = 0$, every curve equals zero. Familiar 1-month, 3-month, 6-month, and 1-year returns are therefore selected points on the same underlying fixed-endpoint curve rather than separate concepts.

We define RTP in terms of the look-back horizon $h$, which provides a convenient parameterization of the historical window. Equivalently, if $t$ denotes the evaluation date, the corresponding historical calendar purchase date is $u = t - h$. For graphical presentation, we generally use the calendar purchase date $u$ on the horizontal axis because actual dates are more directly interpretable for investment decisions and for marking external events. Thus, the mathematical definition is parameterized by look-back horizon, while the graphical display can be read directly in calendar time.

For two investments A and B, a companion difference curve $D_{AB}(h;\, t) = R_A(h;\, t) - R_B(h;\, t)$ can be plotted. Positive values favor A, negative values favor B, and zero crossings identify changes in relative ranking. For more than two investments, one asset can be selected as a reference so that the comparison remains visually simple. When several investments are overlaid, vertical ordering gives the relative realized performance for matched purchase dates. Crossings indicate that the preferred investment depends on the entry date. Persistent separation indicates that one investment produced the larger realized return over a broad range of entry dates.

The common endpoint provides a natural temporal anchor for multi-asset comparison and addresses two related limitations of conventional fixed-start displays. First, even when all investments have sufficiently long histories, a conventional comparison requires the analyst to select a common historical starting date. The resulting comparison is conditional on that choice: different starting dates can produce different relative performance patterns and potentially different conclusions. Second, when investments have different inception dates, imposing a common historical origin may substantially truncate the available histories of older investments, while allowing each investment to begin at its own inception produces returns based on different calendar periods and holding horizons. RTP addresses both issues by reversing the temporal alignment. Rather than fixing one historical origin, it fixes a common calendar endpoint—naturally the present for a current comparison—and allows historical purchase dates to vary over each investment's available history. Section 2.1 illustrates the sensitivity of conventional comparisons to the selected starting date, and Section 2.2 shows how the same endpoint-based construction accommodates investments with unequal inception dates.

### 2.1 Entry-date sensitivity and the choice of starting date

To illustrate the first issue, we compare NVIDIA (NVDA) and Advanced Micro Devices (AMD), two major semiconductor companies competing in high-performance computing, data-center acceleration, and artificial intelligence. Figure 1 presents the conventional fixed-start comparison at several commonly used horizons, which illustrates the central difficulty of a fixed-start comparison. The apparent conclusion depends strongly on the horizon selected. A one-month comparison can favor one stock, while a six-month, one-year, or five-year comparison can produce a very different impression. Each panel is correct for its own starting date, but there is no obvious reason that one of these prespecified dates should be treated as the decisive one. Looking only at these panels therefore does not provide a unified answer to which stock has performed better across the range of plausible historical entry dates.

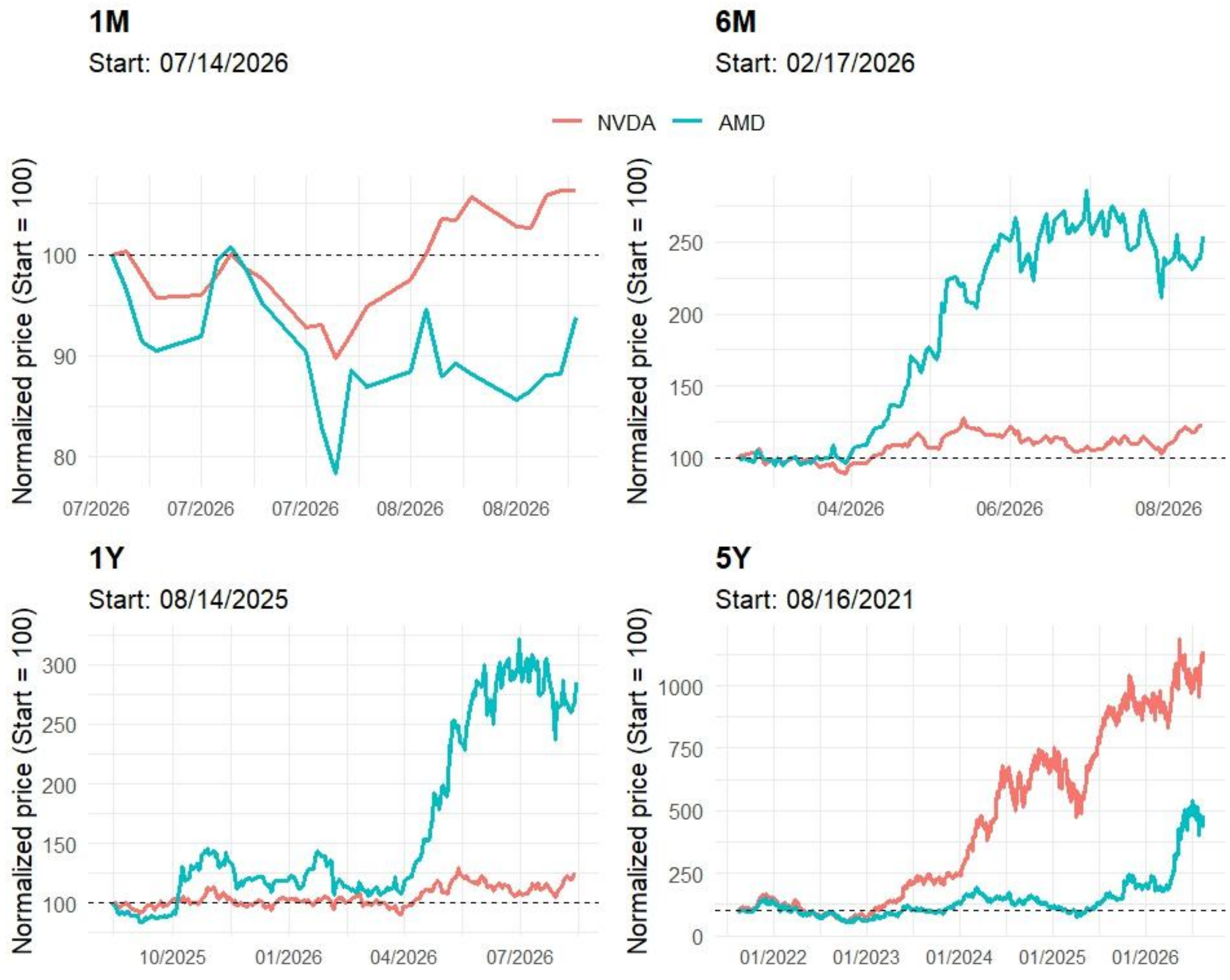

Figure 1. Conventional fixed-start comparisons of NVDA and AMD at selected horizons. Each panel normalizes both stocks to 100 at its own starting date, illustrating how the apparent relative performance changes with the chosen horizon.

The panels also do not show clearly when the relative performance began to change. If the ranking differs between a shorter and a longer horizon, we know that a transition occurred somewhere between those starting dates, but Figure 1 does not identify where the crossover occurs, how persistent the newer pattern is, or whether the separation emerged gradually or abruptly. Recovering that information with the conventional display would require repeatedly changing the starting date and redrawing the comparison. These limitations provide the practical motivation for RTP.

Figure 2 replaces the small set of prespecified horizons with the full Return-to-Present profile. For every historical entry date in the displayed window, the left panel reports the realized return to the same evaluation date. The two curves can therefore be compared continuously rather than only at the few starting dates selected for Figure 1. Their vertical ordering shows which stock produced the larger realized return for a matched entry date, while crossings show where the relative ranking changes.

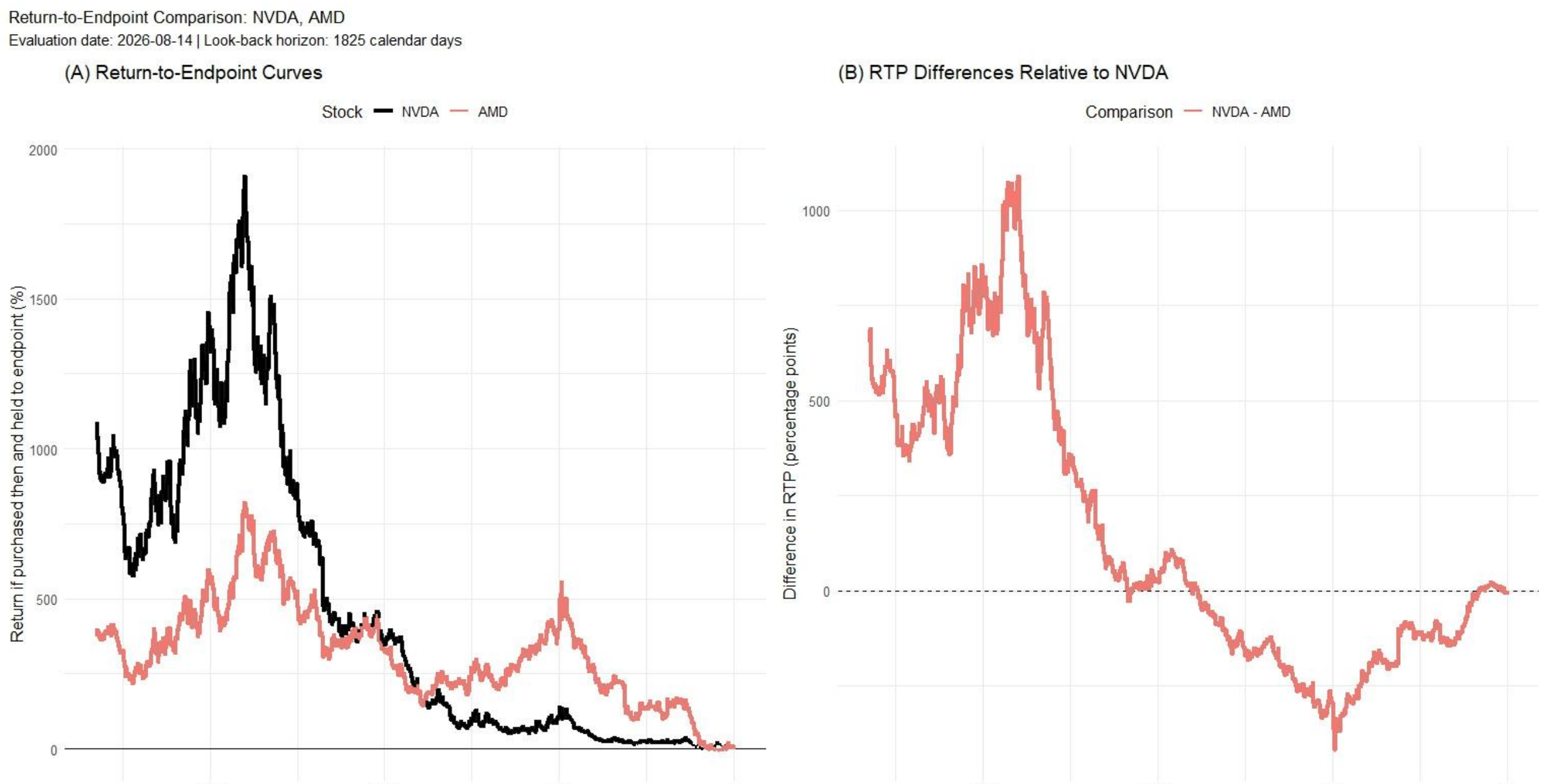

Figure 2. Return-to-Present comparison of NVDA and AMD, with a companion RTP difference curve.

The companion difference curve in the right panel makes the comparison even more direct. Values above zero indicate entry dates for which NVDA has the larger RTP, values below zero indicate entry dates for which AMD has the larger RTP, and zero crossings locate changes in relative ranking. In Figure 2, NVDA substantially outperformed AMD for earlier historical purchase dates, extending roughly three to five years before the evaluation date, with the NVDA-minus-AMD difference reaching several hundred percentage points and exceeding 1,000 percentage points for some entry dates. The relative ordering changes, however, for more recent entry dates. After a crossover at approximately three years before the endpoint, AMD outperformed NVDA over a substantial portion of the subsequent period, with its relative advantage becoming particularly pronounced for some entry dates around 500 days before the evaluation date. The difference then narrows again as the purchase date approaches the endpoint. Thus, the shape, sign, and distance of the difference curve from zero reveal not only which investment performed better, but also when the relative ranking changed and how large and persistent the advantage was. Compared with Figure 1, Figure 2 therefore does not require the reader to select a single historical starting date or mentally reconcile comparisons based on several alternative starting points. Instead, it places the entire fixed-endpoint comparison on one common axis and makes both the timing and magnitude of changes in relative performance directly visible.

## 2.2 Comparing investments with unequal inception dates

A distinct practical difficulty arises when investments have different inception dates. Conventional cumulative-return charts generally require the compared investments to be normalized at a common starting date. When one investment is substantially newer than the others, the comparison must therefore begin no earlier than the inception of the newest investment. Although this produces a valid comparison over the common period, it necessarily discards potentially informative historical performance available for the older investments. In some settings, a desired long-horizon conventional comparison cannot be constructed at all because one or more of the investments did not yet exist at the selected starting date.

Figure 3 illustrates this issue using three AI-focused ETFs with markedly different inception dates. AIQ (Global X Artificial Intelligence & Technology ETF), launched on May 11, 2018, is the oldest of the three and provides broad exposure to companies involved in artificial intelligence and related technologies. CHAT (Roundhill Generative AI & Technology ETF), launched on May 18, 2023, is an actively managed fund focused more specifically on generative AI and related technology companies. AIS (VistaShares Artificial Intelligence Supercycle ETF), launched on December 3, 2024, is the newest of the three and emphasizes the AI infrastructure ecosystem, including high-performance semiconductors, AI data centers, and related applications. Thus, although all three funds provide exposure to the broad AI investment theme, their available historical records differ substantially.

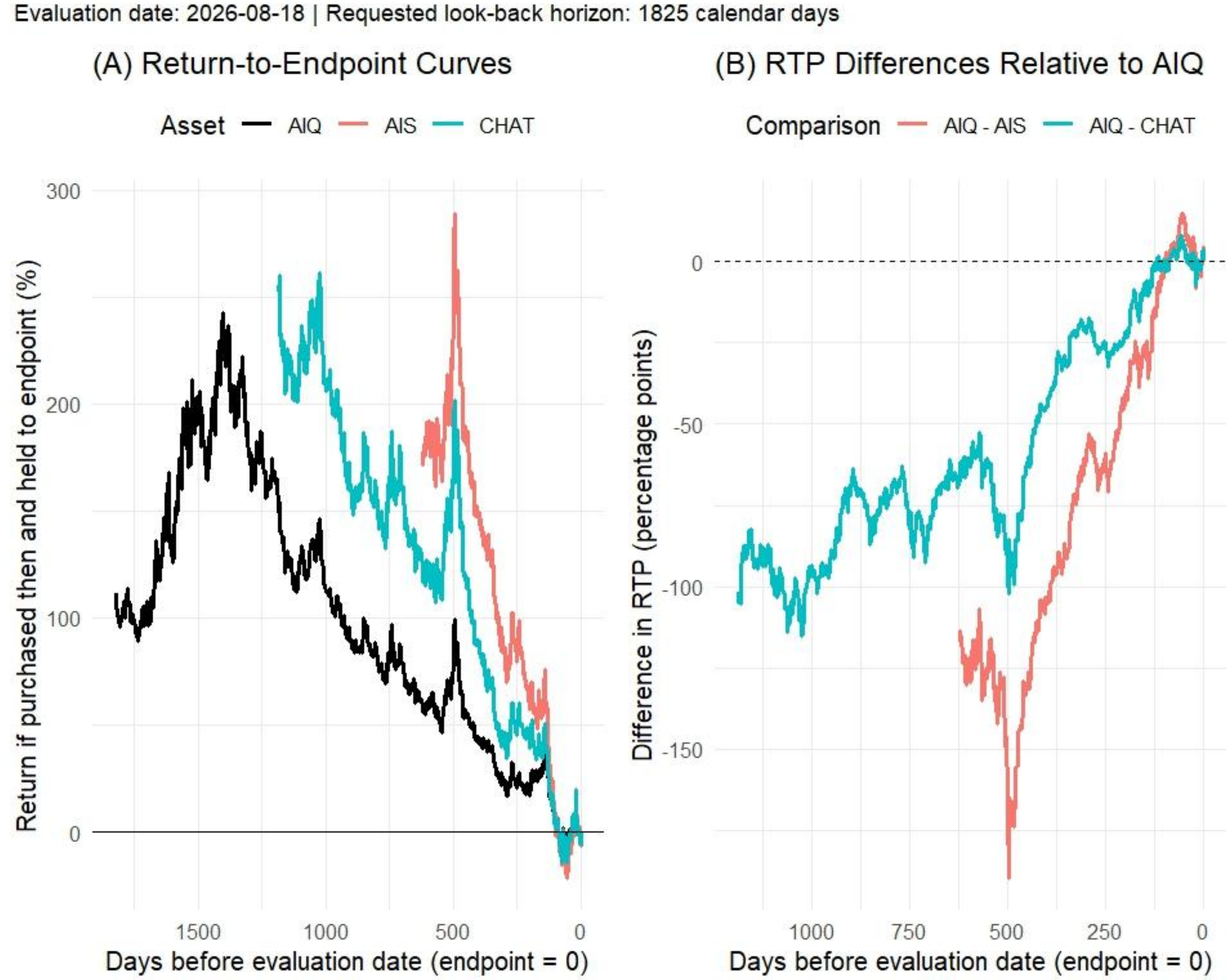

Figure 3. Return-to-Endpoint comparison of AIQ, CHAT, and AIS with unequal inception dates. Each RTP curve retains the full available history of the corresponding ETF while sharing the same evaluation endpoint. Difference curves are defined only over the overlapping histories of the corresponding pairs.

These unequal inception dates create an immediate difficulty for conventional cumulative-return displays. One approach is to enforce a common starting date. A five-year cumulative-return chart beginning in 2021, for example, could include AIQ but not CHAT or AIS because the latter two did not yet exist. To display all three funds with a common origin, the starting date would have to be moved forward to December 3, 2024, the inception date of AIS, thereby discarding more than six years of available history for AIQ and approximately eighteen months of additional history for CHAT. A related limitation arises when financial platforms instead retain each asset's full available history and report cumulative price or total returns from that asset's own inception, often displaying the resulting percentages side by side. Although each return is individually valid, the percentages correspond to different calendar starting dates and holding horizons and therefore are not directly comparable as common-horizon performance measures. This can be particularly problematic for recently launched funds, because a large cumulative return over a short history may appear alongside a smaller return accumulated over a much longer period. Thus, conventional displays face a trade-off: enforcing a common origin discards the longer histories of older investments, whereas retaining asset-specific origins produces cumulative returns measured over incomparable horizons. RTP avoids this ambiguity by anchoring all comparisons to a common calendar endpoint while allowing each asset to contribute its available historical entry dates. Differences in inception dates therefore lead naturally to truncation of the RTP curves rather than to returns calculated over different holding horizons.

This feature is particularly useful for rapidly evolving investment categories such as AI-focused ETFs, in which new products are introduced frequently. The unequal left endpoints in Figure 3 are therefore not missing information to be corrected; they represent the actual differences in the funds' available histories. More generally, the example illustrates why endpoint anchoring is especially useful for comparing multiple investments: the present supplies a common calendar reference even when the investments do not share a common historical origin. RTP preserves the available history of each fund while keeping the comparison anchored to the same evaluation date.

## 3. Illustrations with Real Investment Data

Having defined RTP and used Section 2 to illustrate two central comparison problems—the sensitivity of conventional fixed-start displays to the chosen starting date and the difficulty created by unequal inception dates—we next present two practical applications of the RTP display: recurring retirement-plan allocation and retrospective evaluation of portfolio rotation decisions.

### 3.1 Recurring investment: retirement mutual funds

We first consider a retirement-plan allocation example using three mutual funds available through the Washington University in St. Louis retirement plan administered through TIAA: Vanguard PRIMECAP Fund Admiral Shares (VPMAX), JPMorgan Large Cap Growth Fund Class R6 (JLGMX), and Vanguard Institutional Index Fund Institutional Plus Shares (VIIIX). The three funds provide alternative ways of obtaining substantial exposure to U.S. large-cap equities but differ in investment approach. VPMAX is an actively managed equity fund, JLGMX is an actively managed large-cap growth fund, and VIIIX is a passive index fund designed to track the S&P 500. They therefore represent realistic alternatives that a participant may consider when allocating retirement contributions within the same employer-sponsored plan.

This setting is particularly well suited to RTP because retirement contributions are typically made repeatedly over time rather than as a single investment. Each contribution has a different purchase date, while all accumulated holdings are ultimately evaluated at the same current endpoint. The relevant question is therefore not simply which fund performed best from one arbitrarily selected starting date, but how the funds compare across the many historical dates on which contributions could have been made. By displaying those dates simultaneously, RTP makes the long-run comparative pattern visible in a single graph and can therefore provide useful historical context for retirement-fund selection. The comparison remains descriptive of realized historical performance and is not a forecast of future returns.

Figure 4 shows a persistent long-horizon advantage for VPMAX over both JLGMX and VIIIX across most of the five-year historical entry-date window. Although the magnitude of the advantage varies over time, the broad relative ordering remains remarkably stable.

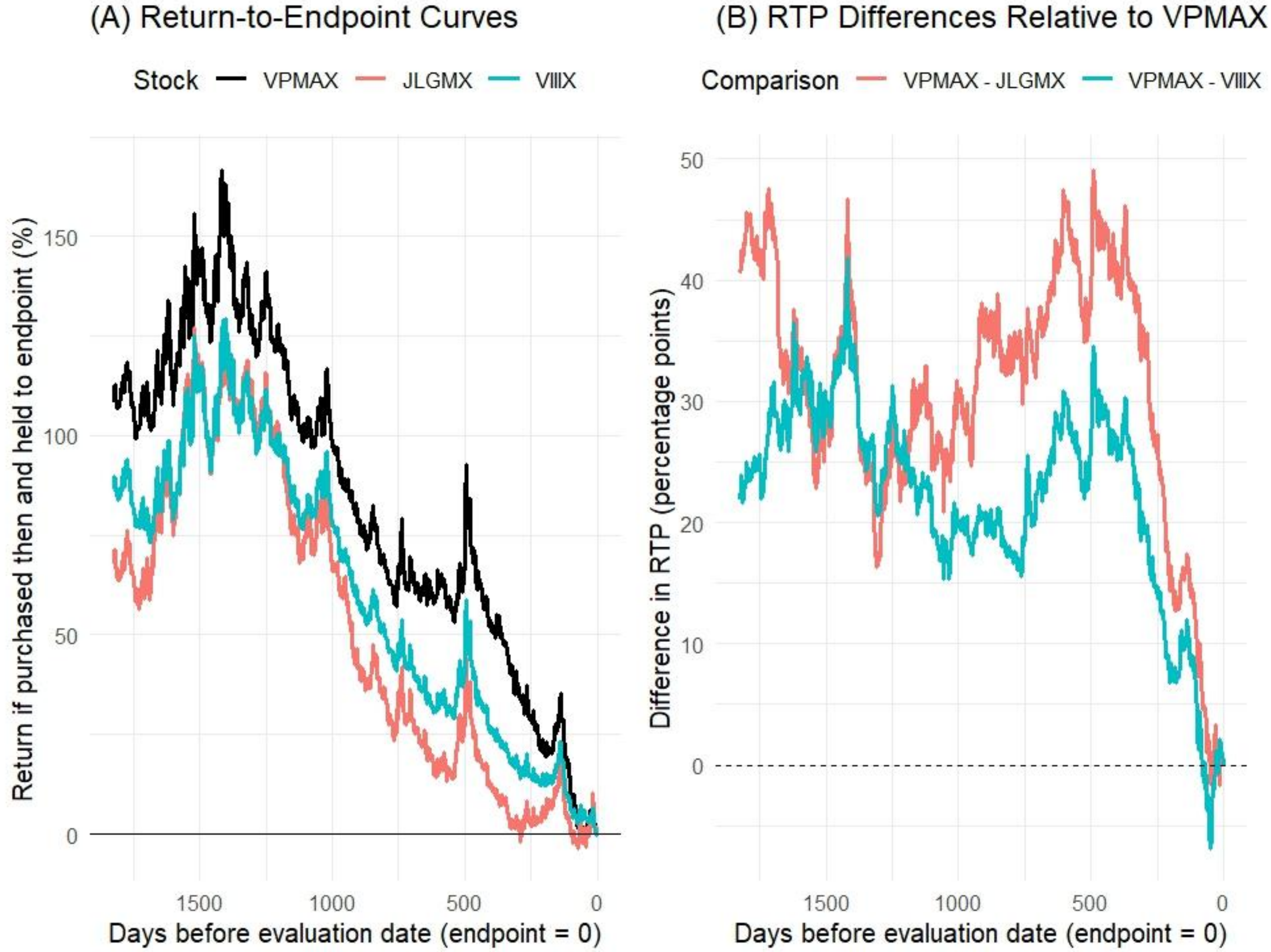

Figure 4. RTP comparison of three retirement mutual funds—VPMAX, JLGMX, and VIIIX—with differences relative to VPMAX.

### 3.2 Evaluating portfolio rotation decisions

RTP can also be used retrospectively to evaluate portfolio rotation decisions. Suppose an investor sells one investment and reallocates the proceeds to another. At a later evaluation date, the question is whether the rotation improved the realized outcome relative to continuing to hold the original investment. Because RTP evaluates historical transaction dates against a common endpoint, the original asset is read at its sale date and the replacement at its actual purchase date. The two dates may be the same or different; the resulting realized returns remain directly comparable because they terminate at the same endpoint.

This feature is especially useful when the replacement purchase is delayed. A conventional fixed-start chart does not directly represent a transaction pair with different starting dates, whereas a single RTP display can show both the effect of the rotation and its sensitivity to the replacement-purchase date without repeatedly resetting the origin or recomputing separate comparisons. The exercise remains retrospective; it is not a trading recommendation or an event study.

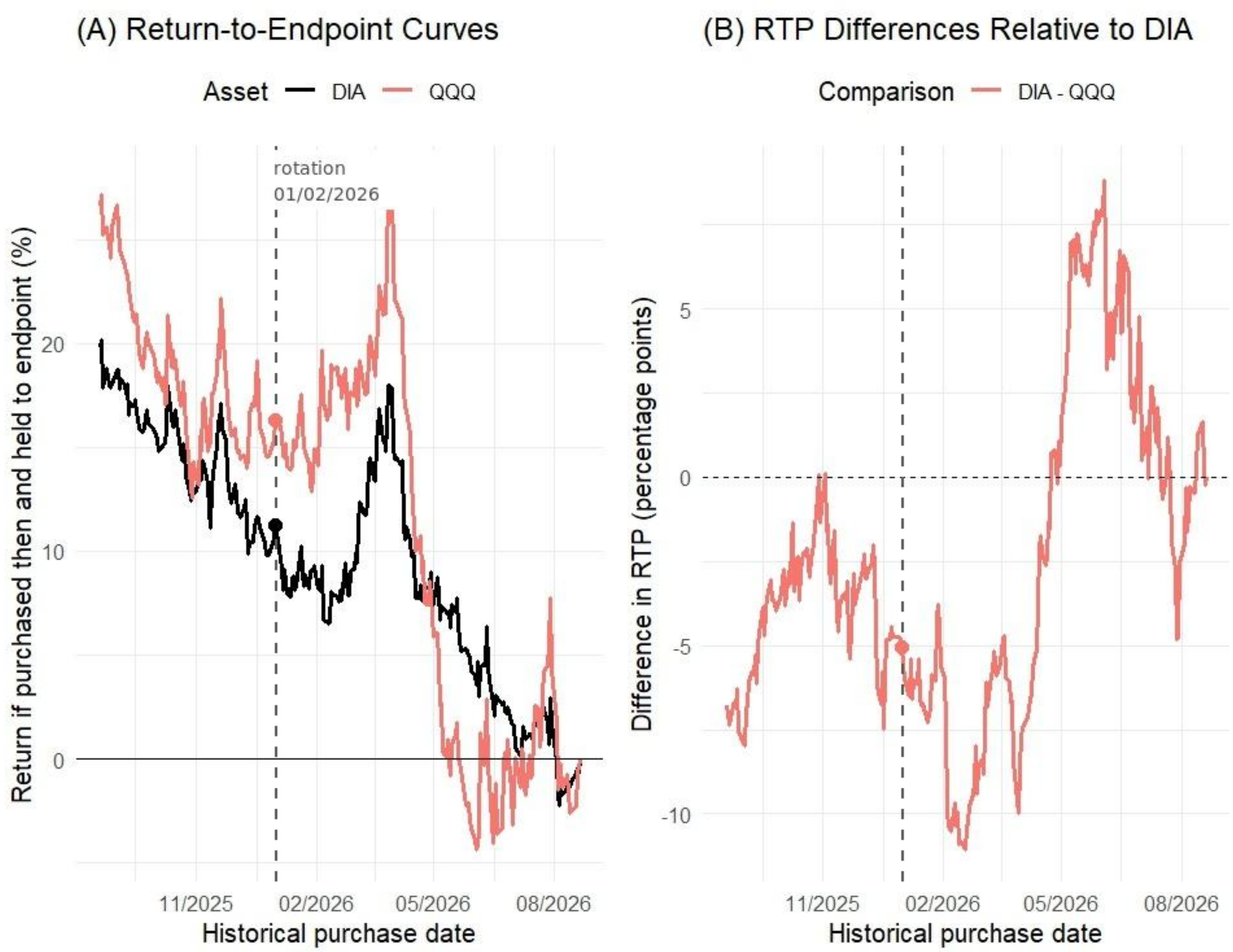

Figure 5. RTP evaluation of a hypothetical reallocation from DIA to QQQ on January 2, 2026, the first trading day of the year, with an August 20, 2026 endpoint. Panel B reports DIA minus QQQ; negative values indicate that QQQ produced the larger realized return. The date is a neutral calendar-based illustration rather than a response to a specific event.

Figure 5 applies this framework to the SPDR Dow Jones Industrial Average ETF Trust (DIA) and the Invesco QQQ ETF (QQQ). DIA tracks a price-weighted portfolio of 30 established U.S. blue-chip companies across several sectors, whereas QQQ tracks 100 of the largest nonfinancial companies listed on Nasdaq and is more heavily oriented toward technology, innovation, and growth. The pair therefore represents a familiar allocation rotation between a more traditional blue-chip portfolio and a more growth-oriented portfolio, while remaining investable at the same level of analysis.

To avoid tying the example to a particular news release or selecting a date because of the subsequent outcome, the chart marks January 2, 2026, the first trading day of the year, as a neutral calendar-based rotation date. The illustration assumes that an investor sold DIA and reinvested in QQQ on that date, then evaluates both alternatives at the common endpoint of August 20, 2026. The marked date is therefore a fixed analytical anchor rather than a claim that a specific event caused the rotation.

At the marked date, QQQ's RTP is approximately five percentage points higher than DIA's; equivalently, the DIA-minus-QQQ difference in Panel B is about -5 percentage points. Under the stated same-day assumption, the hypothetical rotation from DIA to QQQ therefore improved the realized outcome through August 20. The full curves also show why timing matters: QQQ generally has the larger RTP for purchase dates from late 2025 through much of early 2026, while DIA leads for several purchase dates from roughly May through July, with the difference narrowing again near the endpoint.

The same figure also accommodates a staggered rotation. For example, if DIA were sold on January 2 but QQQ were purchased on January 3, the DIA RTP at January 2 can be compared directly with the QQQ RTP at January 3 because both terminate on August 20. Other possible replacement dates can be evaluated in the same way from the same curves. This staggered setting highlights the fixed-endpoint advantage of RTP: each transaction retains its actual date while remaining comparable at the common endpoint. These comparisons remain retrospective, and differences between DIA and QQQ reflect their distinct holdings, sector exposures, concentration, and weighting methods; they should not be interpreted as a pure sector effect, a causal estimate, or evidence that the rotation was prospectively optimal.

## 4. Discussion

The principal contribution of RTP is visual rather than computational. The underlying returns are familiar, but RTP organizes them as a continuous fixed-endpoint performance profile over historical purchase dates, making patterns such as persistence, separation, crossings, and timing dependence directly visible. The RTP curve is therefore the primary output, while descriptive summaries such as the proportion of historical entry dates for which one investment has the larger RTP or the average RTP difference may be reported secondarily. The same representation is not restricted to individual stocks; it can be applied to any comparable investment price series, including mutual funds, ETFs, and market or sector indices, provided that the common-endpoint interpretation is substantively meaningful.

A practical use of this representation is the comparison of two otherwise plausible investments when an investor must decide which one has shown the stronger realized historical performance. The relevant criterion need not be that one investment dominates the other at every possible historical purchase date, which would often be unrealistic. Instead, the RTP curves can show whether one investment exhibits broad or predominant historical dominance—that is, whether it produces the larger realized return to the common endpoint for a substantial majority of historical purchase dates. The proportion of entry dates favoring each investment, together with the magnitude and persistence of the RTP differences, can then serve as a descriptive summary of the comparison. This does not establish future superiority, but it provides a direct historical criterion for choosing between investments that may otherwise appear comparable.

This historical-comparison perspective is also related to the familiar concept of momentum [1,2]. Conventional momentum measures typically summarize past performance over one or a small number of prespecified look-back horizons. RTP extends this descriptive perspective graphically by displaying realized past performance continuously across the full range of historical look-back horizons. It can therefore be interpreted as a continuous multi-horizon profile of momentum-like realized performance, showing not only the strength of past performance but also how relative rankings change as the look-back horizon varies. An important distinction, however, is that neighboring RTP values are strongly dependent because they share the same endpoint and largely overlapping price histories. Consequently, summaries such as the proportion of historical entry dates favoring one investment or the number of curve crossings describe the displayed historical performance profile; they should not be interpreted as independent evidence or as probabilities of future outperformance. RTP is therefore not proposed as a new momentum factor or trading signal, and whether the historical patterns displayed by RTP have predictive value for future returns is a separate empirical question.

RTP is particularly informative for peer comparisons within the same industry or sector. When firms share broad macroeconomic and sector-level exposures, common movements tend to affect their curves in similar directions, so persistent separation, crossings, or abrupt changes in a pairwise difference curve can make firm-specific divergence or competitive shifts easier to identify. The semiconductor comparison in Figure 2 illustrates this use. A common-sector setting does not eliminate differences in size, business mix, leverage, or risk, however, and visual separation should not be interpreted as causal evidence.

The choice between RTP and a conventional forward-time display should depend on the question being asked. RTP is particularly useful when the analysis is anchored to a common endpoint and the historical purchase or entry date is allowed to vary, as in comparisons of historical dominance, recurring investment, unequal inception dates, or retrospective rotation decisions. In these settings, a conventional cumulative-return comparison conditions on a selected starting date, and the apparent relative performance may change when that date is changed; recovering the full entry-date pattern would require repeatedly resetting the origin. RTP instead holds the endpoint fixed and displays the continuum of historical entry dates in a single representation. Conversely, RTP is not intended to replace conventional forward-time plots when the primary question concerns how two series move together through calendar time, including temporal co-movement or possible lead-lag patterns. For such questions, ordinary forward trajectories may be more natural and equally or more informative. Thus, the principal advantage of RTP is not universal superiority over forward plots, but its direct representation of common-endpoint, entry-date-dependent realized performance.

RTP remains conditional on the selected evaluation endpoint. The present provides a natural default for current investment comparison, whereas an alternative historical endpoint may be appropriate when motivated by a specific market, company, or policy event. Because a large price movement near the endpoint affects returns over many look-back horizons, exceptional recent episodes can substantially influence the curve. Alternative endpoints should therefore be substantively motivated rather than selected to favor a particular result, and sensitivity to exceptional endpoint-adjacent periods warrants further study.

Intermediate events within the historical window require a different interpretive principle. Historical intervals should not be removed mechanically simply because they appear abnormal or coincide with unusual news. If an event produces a persistent change in valuation or market regime, its effect is carried forward in the subsequent realized price path and is therefore part of the historical performance that RTP is designed to display; removing the interval would also remove information embedded in later outcomes. By contrast, a transitory shock—such as false or rapidly corrected news followed by a prompt price recovery—typically affects RTP only for purchase dates near that episode and has limited influence on the longer-horizon profile. Accordingly, any exclusion or alternative treatment of an intermediate period should be based on the substantive persistence of its effect, not merely on the unusual appearance of the event or the associated price movement.

We view RTP as a simple communication and decision-support visualization that complements, rather than replaces, conventional price and cumulative-return charts, trailing-return tables, and momentum summaries. A longer-term objective is to incorporate RTP into brokerage, financial-information, and retirement-investment platforms as an interactive display, allowing investors to compare assets at a common endpoint, inspect specific historical purchase dates and returns, mark actual purchase, sale, or switching dates, and explore nearby alternative transaction timings. Such implementation would make continuous fixed-endpoint comparison readily accessible to individual investors during real investment and allocation decisions.

## 5. Implementation and Reproducibility

A reproducible R implementation can take as inputs a ticker list, look-back horizon, optional reference asset, and evaluation date. The default evaluation date is the current date; users may also specify a historical endpoint. Adjusted prices are used so that the return calculation reflects distributions and corporate actions represented in the adjusted series. The software can produce the RTP overlay, reference-based difference curves, and concise descriptive summaries. Although the calculation is parameterized by look-back horizon, the horizontal axis can be displayed using the corresponding calendar purchase dates to facilitate interpretation and event annotation.

## References

[1] Jegadeesh N, Titman S. Returns to Buying Winners and Selling Losers: Implications for Stock Market Efficiency. Journal of Finance. 1993;48(1):65-91. doi:10.1111/j.1540-6261.1993.tb04702.x.

[2] Moskowitz TJ, Ooi YH, Pedersen LH. Time Series Momentum. Journal of Financial Economics. 2012;104(2):228-250. doi:10.1016/j.jfineco.2011.11.003.